\documentclass[10pt,conference]{IEEEtran}
\IEEEoverridecommandlockouts
\usepackage{cite}
\usepackage{amsmath,amssymb,amsfonts}
\usepackage{algorithmic}
\usepackage{graphicx}
\usepackage{textcomp}
\usepackage{xcolor}
\usepackage{url}
\def\BibTeX{{\rm B\kern-.05em{\sc i\kern-.025em b}\kern-.08em
    T\kern-.1667em\lower.7ex\hbox{E}\kern-.125emX}}

\usepackage{preamble}
\begin{document}
\title{T2T: Captioning Smartphone Activities\\Using Mobile Traffic}
 
\author{ 
  Jiyu Liu$^{\dagger}$,
  Yong Huang$^{*^{\dagger}}$,
  Yanzhao Lu$^{\dagger}$,
  Yun Tie$^{\dagger}$,
  Wanqing Tu$^{\S}$ \\

  $^{\dagger}$School of Cyber Science and Engineering, Zhengzhou University, Zhengzhou, China \\
  $^{\S}$Department of Computer Science, Durham University, Durham, United Kingdom \\
  
  Email: \{jiyuliu, yanzhaolu\}@gs.zzu.edu.cn, \{yonghuang, ieytie\}@zzu.edu.cn,
  wanqing.tu@durham.ac.uk

  \thanks{\textsuperscript{*}The corresponding author is Yong Huang (yonghuang@zzu.edu.cn).}
}

\maketitle

\begin{abstract}
This paper studies the creation of textual descriptions of user activities and interactions on smartphones. 
Our approach of referring to encrypted mobile traffic exceeds traditional smartphone activity classification methods in terms of model scalability and output readability.
The paper addresses two obstacles to the realization of this idea: the semantic gap between traffic features and smartphone activity captions, and the lack of textually annotated traffic data.
To overcome these challenges, we introduce a novel smartphone activity captioning system, called T2T (Traffic-to-Text).
T2T consists of a flow feature encoder that converts low-level traffic characteristics into meaningful latent features and a caption decoder to yield readable transcripts of smartphone activities.
In addition, T2T achieves the automatic textual annotation of mobile traffic by feeding synchronized screen capture videos into the Qwen-VL-Max vision–language model, and proposing multi-stage losses for effective cross-model training.
We evaluate T2T on 40,000 traffic-description pairs collected in two real-world environments, involving 8 smartphone users and 20 mobile apps.
T2T achieves a BLEU-4 score of 58.1, a METEOR score of 38.3, a ROUGE-L score of 70.5, and a CIDEr score of 108.7.
The quantitative and qualitative analyses show that T2T can generate semantically accurate captions that are comparable to the vision–language model.
\end{abstract}

\begin{IEEEkeywords}
Smartphone Activity, Captioning, Network Traffic Analysis
\end{IEEEkeywords}

\section{Introduction}
\begin{figure}[t]
    \centering
    \includegraphics[width=0.9\linewidth]{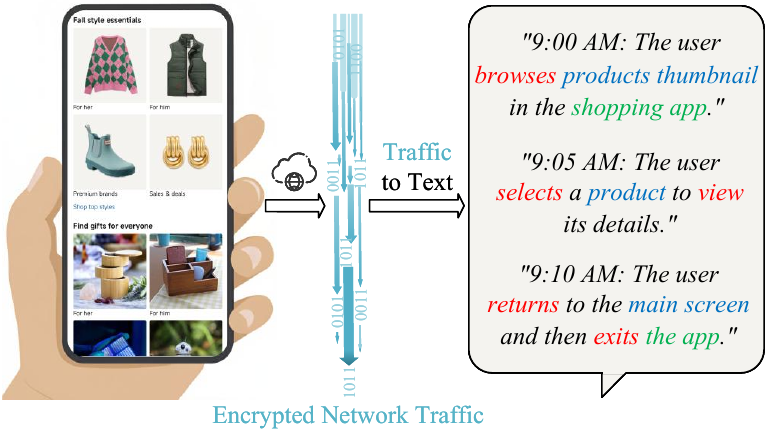}
    \caption{Illustration of a ``Traffic-to-Text'' system.}
    \label{fig:illustration}
\end{figure}

Captioning, typically referring to the process of adding text to visual content (e.g., an image, a video), plays a vital role in enabling computer vision and natural language processing, enhancing the interpretability and accessibility of visual content\cite{alb2023attention}.
This paper brings these features into the network traffic analysis domain and focuses on smartphone activity captioning using mobile traffic. 
This goal necessitates a ``Traffic-to-Text'' system that receives network traffic from smartphones and generates textual descriptions of ongoing user operating behaviors on smartphones, as depicted in Fig.~\ref{fig:illustration}.
Such a system can generate chronological narratives of user-smartphone interactions, without accessing on-device smartphone usage logs~\cite{song2025predicting} as existing methods require. 
These capabilities of the system will enable many intriguing applications, such as monitoring children's nighttime mobile phone usage on home Wi-Fi networks, and hyper-personalized mobile experiences or advertisements provided by network service providers.

Captioning smartphone activities significantly differs from the existing task of smartphone activity classification~\cite{ zhuo2020real, yang2024eavesdropping}.
First, captioning is a generative task that outputs combinations of words from a large-scale vocabulary, rather than from a limited number of predefined apps and in-app actions.
Second, captioning can generate narrative-style descriptions of user-smartphone interactive events consisting of apps being used, action sequences, in-app functions, and other detailed information.
These advantages make smartphone activity captioning more scalable, readable, and attractive than smartphone activity classification.

However, using mobile traffic analysis to caption user activities also introduces new challenges.
The first challenge is \textit{the semantic gap between traffic features and captions.}
In the domain of network traffic analysis, encrypted network flows are typically represented by numerical attributes, such as packet directions, sizes, and rates. These low-level attributes have no direct connection to user activities or interactions, posing a challenge for the derivation of meaningful inferences for semantically accurate captions.
The second challenge is \textit{the lack of textually annotated traffic data.}
Training a captioning system generally requires a large number of textually labeled samples, although there are currently no such datasets for network traffic.
However, manually annotating mobile traffic with natural language scripts would be a daunting task.

In this paper, we propose T2T (Traffic-to-Text), a smartphone activity captioning system that addresses the above two challenges and hence generates semantically accurate narrative captions describing the activity of smartphone users.
To bridge the semantic gap, T2T first devises a novel flow feature encoder that integrates app type information into the latent features through dynamic feature modulation and discovers representative flow patterns in each type of smartphone activities via prototype learning.
Then, a caption decoder, i.e., an attention-based long short-term memory (LSTM) network, is developed to translate the learned features into natural language descriptions.
To effectively train T2T, we propose a cross-modal annotation and training scheme. 
It exploits Qwen-VL-Max~\cite{bai2023qwen}, a state-of-the-art multimodal large language model, to automatically generate ground-truth captions based on synchronized video clips from smartphone screens.
Then, multi-stage losses are leveraged to transfer visual knowledge of smartphone activities to T2T. The source code is available at \url{https://github.com/PhyGroup/T2T}.

The contributions of our work are summarized as below.
\begin{itemize}
\item T2T is among the first systems to generate accurate and narrative captions from analyzing encrypted network traffic. It proposes a novel method to understand user behaviors in cyberspace.

\item T2T is an automated caption system that derives semantic captions from smartphone traffic instead of manually prepared datasets. This automation greatly enhances the efficiency of model training.

\item T2T is evaluated with 40,000 traffic-description pairs collected from 8 smartphone users and 20 mobile applications. Our evaluations show that the captioning performance of T2T is comparable to that of a vision-language model.
\end{itemize}


\section{Related Work}
\textbf{Captioning.}
Captioning aims to generate natural language descriptions from non-linguistic data. 
Image captioning is the earliest line in this domain~\cite{zhang2024interactive}, typically relying on region-based or grid-based visual representations to ground semantic descriptions.
Video captioning extends this task to dynamic visual sequences, requiring modeling of temporal context and event evolution across frames~\cite{HMN}, which facilitates applications such as human-robot interaction, video indexing, and describing visual content for the blind.
Beyond visual modalities, audio captioning has recently emerged~\cite{xu2024audiocap}, which describes acoustic environments by interpreting non-speech sounds such as footsteps, vehicle horns, or kitchen noises.
In addition, radio signals are exploited to caption in-home daily life~\cite{fan2020home}, offering a privacy-preserving way to monitor and describe human activities without visual or audio recording.
In this paper, we focus on captioning smartphone activities using encrypted network traffic.

\textbf{Smartphone Activity Classification.}
Smartphone activity classification is used to identify user interactions with their smartphones~\cite{zhuo2020real}. 
The existing methods mainly depends on screen recording~\cite{zhang2024screen}, device usage logs~\cite{song2025predicting}, or on-device sensors~\cite{sensor2024dcapsnet} to obtain detailed information, while facing serious privacy concerns. 
As an alternative, some studies classify smartphone activities using network traffic. 
Feature-based approaches exploit statistical characteristics to achieve high performance with expert-engineered features~\cite{li2022foap, huang2025TMC}. 
Deep learning methods, including recurrent neural networks (RNNs)~\cite{RNN} and convolutional neural networks (CNNs)~\cite{chen2025cdnet}, are used to extract features from raw traffic. 
Although high classification performance is achieved, they are limited to specific mobile applications or predefined in-app labels. 
In contrast, T2T extends the structured label representation to user-friendly natural language descriptions, providing richer and more interpretable insights into user-smartphone interactions.

\section{System Design}
\label{sec:method}

\subsection{System Overview}
\begin{figure*}[ht]
    \centering
    \includegraphics[width=0.98\linewidth]{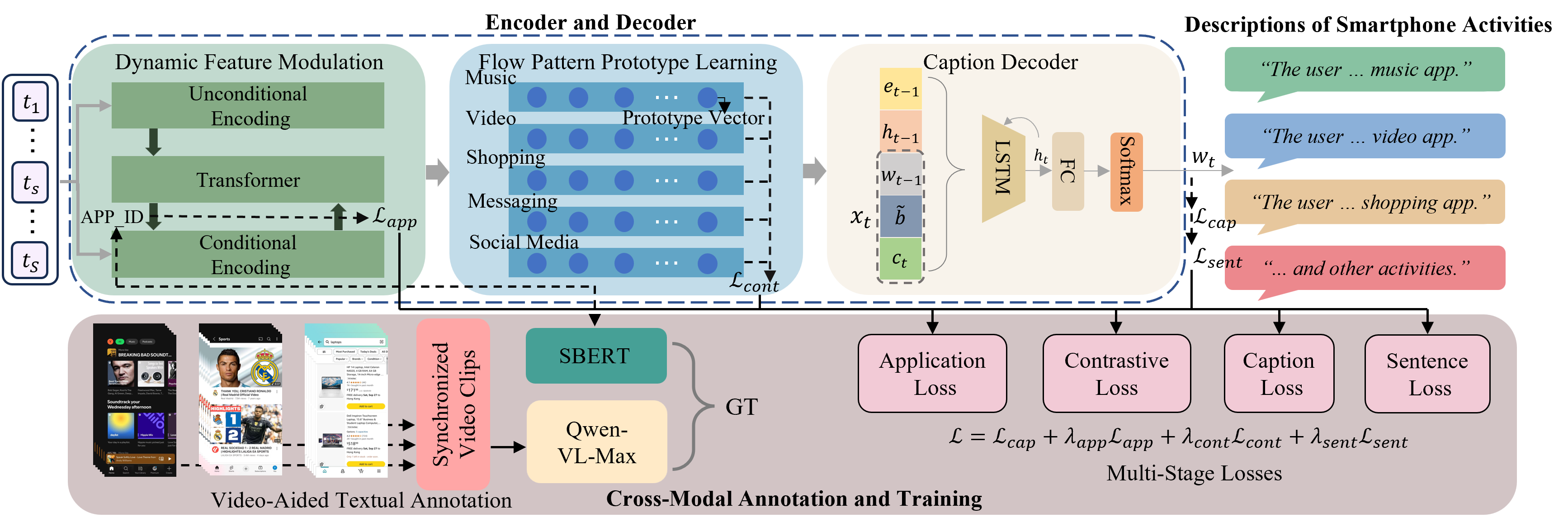}
    \caption{Overview of T2T. It contains a flow feature encoder, a caption decoder, and a cross-modal annotation and training scheme.}
    \label{fig:framework}
\end{figure*}

This paper introduces T2T, a ``Traffic-to-Text" system that converts encrypted network traffic into fine-grained textual descriptions of users' activities and interactions with their smartphones.   
T2T can be strategically positioned at key network junctures to capture and process all inbound and outbound traffic effectively. 
Typically, it can be deployed either at the network gateway within a smart home or enterprise environment or at the network service provider side.
Moreover, we assume that the network operator can isolate the traffic from mobile devices through network management information such as IP or DHCP records, and feed it into T2T.
As shown in Fig.~\ref{fig:framework}, T2T consists of a flow feature encoder, a caption decoder, and a cross-modal annotation and training scheme.
We will elaborate on these components in the following subsections. 

\subsection{Flow Feature Encoder and Caption Decoder}
\label{sec:dynamic_feature_modulation}

We define the network flow as a sequence of packets corresponding to a socket-to-socket communication, which can be identified by a unique combination of source and destination addresses and port numbers, together with transport protocols.  
Each flow encodes $D$ flow-level attributes and is represented as a feature vector $\mathbf{t}_s \in \mathbb{R}^{1 \times D}$.
We leverage a series of $S$ consecutive flows as a basic input, which can be represented as a feature sequence $\mathbf{T} = \{\mathbf{t}_1, \cdots, \mathbf{t}_s, \cdots, \mathbf{t}_S\} \in \mathbb{R}^{S \times D}$, ordered by the timestamp of the earliest packet in each flow.
Because a feature sequence $\mathbf{T}$ contains flow-level numerical attributes of encrypted network traffic, the first step of T2T is to transform these values into meaningful latent features.
For this purpose, we devise a flow feature encoder that characterizes flow patterns in each type of mobile apps.

\textbf{Dynamic Feature Modulation.}
Since user-smartphone interactions are predictable within a type of mobile apps~\cite{yang2024eavesdropping}, we propose a two-stage feature modulation process to encode app type information into learned feature representation. 

In the first stage, unconditional encoding is performed to infer the application type of the feature sequence $\mathbf{T}$ to guide subsequent feature modulation. 
The feature vectors in $\mathbf{T}$ are individually projected into a higher-dimensional embedding space via a linear transformation to obtain \( \mathbf{T}' = \{\mathbf{t}'_1;\cdots;\mathbf{t}'_s;\cdots;\mathbf{t}'_S\} \), where \( \mathbf{t}'_s \in \mathbb{R}^{1\times H} \) and \( H \) is the dimension of the embedding space.
A transformer, with sinusoidal positional encoding on the vector index \( s \), processes \( \mathbf{T}' \) to capture underlying temporal characteristics as $\mathbf{T}'' = \text{Transformer}(\text{PosEncode}(\mathbf{T}'))$, where \( \mathbf{T}'' = \{\mathbf{t}''_1;\cdots;\mathbf{t}''_s;\cdots;\mathbf{t}''_S\} \) is the unconditional encoded feature sequence and \( \mathbf{t}''_s \in \mathbb{R}^{1\times H} \) represents an unconditional encoded feature vector. 
After that, the mean pooling is applied on $\mathbf{T}''$, followed by a fully-connected (FC) layer and a softmax layer to infer the corresponding app type.
The above process can be expressed as \begin{equation}
\mathbf{p} = \text{Softmax}(\text{FC}(\text{Mean}(\mathbf{T}''))),
\label{eq:app_type_pred}
\end{equation}
where \( \mathbf{p} \in [0,1]^{1\times K} \) is the estimated probability distribution over \( K \) app types.
The index of the predicted app type $\hat{k}$ can be obtained as \( \hat{k} = \arg\max(\mathbf{p}) \).
According to the work~\cite{yang2024eavesdropping}, we categorize mobile apps into five types, including music, video, shopping, messaging, and social media.

In the second stage, the conditional encoding is leveraged to enhance the representation capability of $\mathbf{T}''$ by incorporating the semantic information of the predicted app type into it.
To achieve this, we employ feature-wise linear modulation (FiLM)~\cite{perez2018film}, a technique that dynamically scales and shifts latent feature representations.
First, we maintain an app type embedding matrix \( \mathbf{E} \in \mathbb{R}^{K \times L} \), where \( L \) is the embedding dimension.
In $\mathbf{E}$, each row is a semantic embedding vector, obtained by feeding one of the five type labels, i.e., ``video", ``music", ``shopping", ``messaging", ``social media", into Sentence-BERT (SBERT)~\cite{2019SentenceBert}.
In this way, two independent FC layers, \( \text{FC}_{\mathbf{\gamma}} \) and \( \text{FC}_{\mathbf{\beta}} \), are used to generate the scaling parameter \( \mathbf{\gamma} \) and bias parameter \( \mathbf{\beta} \) respectively as
\begin{equation}
\begin{aligned}
\mathbf{\gamma} = \text{FC}_{\mathbf{\gamma}}(\mathbf{E}[\hat{k}]) \;\; \text{and} \;\; \mathbf{\beta} = \text{FC}_{\mathbf{\beta}}(\mathbf{E}[\hat{k}]),
\end{aligned}
\label{eq:film_params}
\end{equation}
where \( \mathbf{\gamma}, \mathbf{\beta} \in \mathbb{R}^{1 \times H} \) and $\mathbf{E}[\hat{k}]$ indicates the $\hat{k}$-th row of $\mathbf{E}$.
Next, the two parameters are applied to modulate each feature vector \( \mathbf{t}''_s \) in \( \mathbf{T}'' \) as
\begin{equation}
\tilde{\mathbf{t}}_s = \mathbf{\gamma} \odot \mathbf{t}''_s + \mathbf{\beta},
\label{eq:film_modulation}
\end{equation}
where \( \odot \) denotes element-wise multiplication.
After that, the modulated feature sequence \( \tilde{\mathbf{T}} = \{\tilde{\mathbf{t}}_1;\cdots;\tilde{\mathbf{t}}_s;\cdots;\tilde{\mathbf{t}}_S\} \) is obtained.
Finally, the transformer with sinusoidal positional encoding is used again to obtain the conditional encoded feature sequence $ \mathbf{F} = \left\{ \mathbf{f}_1;\cdots;\mathbf{f}_s;\cdots;\mathbf{f}_S\right\} \in \mathbb{R}^{S \times H}$ as $\mathbf{F} = \text{Transformer}(\text{PosEncode}(\tilde{\mathbf{T}}))$, where \( \mathbf{f}_s \in \mathbb{R}^{1 \times H} \) denotes a conditional encoded feature vector.
Based on $\mathbf{F}$, the mean pooling is applied to yield a global feature vector \( \mathbf{f}_{\text{global}} \in \mathbb{R}^{1 \times H} \) as $\mathbf{f}_{\text{global}} = \text{Mean} (\mathbf{F})$.

\textbf{Flow Pattern Prototype Learning.}
Traffic patterns can reflect in-app actions, and similar patterns may correspond to distinct actions in different app types. 
To characterize in-app actions, we introduce a flow pattern prototype learning layer. 
For each \( k \in \{0, 1, \dots, K-1\} \), this layer models diverse network flow patterns within each app type using \( M \) learnable prototypes \( \mathbf{P}_k = \{\mathbf{p}_{k,1};\cdots;\mathbf{p}_{k,m};\cdots;\mathbf{p}_{k, M}\} \in \mathbb{R}^{M \times H} \), where a prototype is a learnable reference vector in the representation space and encodes a representative traffic pattern rather than a specific sample. 
These prototypes are initialized with the Xavier initialization and refined during training to align with traffic characteristics. 
Given the global feature vector \( \mathbf{f}_{\text{global}} \) and the predicted app type \( \hat{k} \), the attention weight of $m$-th prototype is computed as 
\begin{equation}
\alpha_m = \frac{\exp(\mathbf{f}_{\text{global}} \cdot \mathbf{p}_{\hat{k},m} / \sqrt{H})}{\sum_{m=1}^M \exp(\mathbf{f}_{\text{global}} \cdot \mathbf{p}_{\hat{k},m} / \sqrt{H})},
\label{eq:proto_attn}
\end{equation}
where \(\alpha_m\) represents the correlation between the global traffic representation \( \mathbf{f} \) and the \( m \)-th prototype \( \mathbf{p}_{\hat{k},m} \). 
In this way, a set of attention weights can be obtained as \( \{\alpha_1,\cdots,\alpha_m, \cdots, \alpha_M\} \).
Then, these weights are aggregated to form a pattern embedding \( \mathbf{b} \in \mathbb{R}^{1 \times H} \), representing a weighted prototype combination, which is given by 
\begin{equation}
\mathbf{b} = \sum_{m=1}^M \alpha_m \mathbf{p}_{\hat{k},m}.
\label{eq:proto_embedding}
\end{equation}
After that, the prototype representation $\mathbf{b}$ and the global feature vector $\mathbf{f}_{\text{global}}$ are added and input into a normalization layer as $\mathbf{b}' = \text{Norm}(\mathbf{b} + \mathbf{f}_{\text{global}})$. Next, we refined $\mathbf{b}'$ by a weighted integration with $\mathbf{f}_{\text{global}}$ using a learnable parameter $\alpha \in [0, 1]$, followed by a feed-forward layer, another normalization layer, and a dropout layer, yielding the final flow pattern embedding $\tilde{\mathbf{b}} = \text{Dropout}\left( \text{LayerNorm}\left( \text{FC}\left( \alpha \mathbf{b}' + (1 - \alpha) \mathbf{f}_{\text{global}} \right) \right) \right)$ where $\tilde{\mathbf{b}} \in \mathbb{R}^{1 \times H'}$ and $H'$ is the embedding dimension.

\textbf{Caption Decoder.}
The caption decoder proceeds to translate the pattern embedding $\tilde{\mathbf{b}}$ and the conditional encoded feature sequence $\mathbf{F}$ into natural language descriptions. 
To achieve this goal, we design an attention-based LSTM network that integrates the previous word embedding $\mathbf{w}_{t-1} \in \mathbb{R}^{1 \times L}$, the prior hidden state $\mathbf{h}_{t-1} \in \mathbb{R}^{1 \times H}$, the prior cell state $\mathbf{e}_{t-1} \in \mathbb{R}^{1 \times H}$, along with $\mathbf{F}$ and $\tilde{\mathbf{b}}$ at each time step $t$. Specifically, an attention mechanism computes a context vector $\mathbf{c}_t \in \mathbb{R}^{1 \times H}$ by dynamically weighting the encoded features in $\mathbf{F}$ based on $\mathbf{h}_{t-1}$ as $\phi_{t,s} = \text{Softmax}(\mathbf{r}^\top \tanh(\mathbf{u} \odot \mathbf{h}_{t-1} + \mathbf{v} \odot \mathbf{f}_s + \mathbf{o}))$ and $\mathbf{c}_t = \sum_{s=1}^S \phi_{t,s} \cdot \mathbf{f}_s$, where $\mathbf{r}, \mathbf{u}, \mathbf{v}, \mathbf{o} \in \mathbb{R}^{1 \times H}$ are learnable parameters that encourage the decoder to focus on traffic-related features. 
The LSTM input is then formed by concatenating $\mathbf{w}_{t-1}$, $\mathbf{c}_t$, and $\tilde{\mathbf{b}}$ as $\mathbf{x}_t = [\mathbf{w}_{t-1}; \mathbf{c}_{t}; \tilde{\mathbf{b}}] \in \mathbb{R}^{L+H+H'}$. 
The hidden and cell states are updated as 
\begin{equation}
\left\{ \mathbf{h}_t; \mathbf{e}_t \right\} = \text{LSTM}(\mathbf{x}_t, \mathbf{h}_{t-1}, \mathbf{e}_{t-1}),
\label{eq:lstm_update}
\end{equation}
and the next-word probability distribution $\mathbf{q}_t \in [0,1]^{1\times V}$ over a vocabulary of size $V$ is predicted as $\mathbf{q}_t = \text{Softmax}(\text{FC}(\mathbf{h}_t))$. Finally, the current word is selected by $w_t = \arg\max \mathbf{q}_t$.

\subsection{Cross-Modal Annotation and Training}
\label{sec:training}

Currently, no existing dataset pairs smartphone network traffic with corresponding captions.
While captioning systems typically require large-scale labeled data, manual annotation of mobile traffic is prohibitively labor-intensive.
To address this challenge, we propose a cross-modal annotation and training scheme by leveraging Qwen-VL-Max~\cite{bai2023qwen} to automatically generate textual descriptions, and employing multi-stage losses to transfer visual knowledge of smartphone activities to T2T.

\textbf{Video-Aided Textual Annotation.}
Let us denote $\mathcal{T} =\left\{\mathbf{T}_{i} \right\}^{I}_{i=1}$ as the feature sequences extracted from the collected traffic traces.
During traffic collection, we activate screen recording on targeted smartphones used by volunteers to obtain video clips of user-smartphone interactions, which are fed into Qwen-VL-Max to generate activities transcripts, yielding a synchronized caption set $\mathcal{W} = \left\{\mathbf{w}_{i} \right\}^{I}_{i=1}$ aligned with $\mathbf{T}_{i}$ by timestamps.
In addition, the type of foreground app is also recorded when collecting each traffic trace, forming the app type label set $\mathcal{K} =\left\{k_{i} \right\}^{I}_{i=1}$.
Finally, we combine the three sets $\mathcal{T}$, $\mathcal{W}$, and $\mathcal{K}$ into a training dataset as $\mathcal{D}=\left\{(\mathbf{T}_{i}, \mathbf{w}_{i}, k_{i}) \right\}^{I}_{i=1}$.

\textbf{Multi-Stage Losses.} 
We introduce an app loss $\mathcal{L}_{\text{app}}$, a contrastive loss $\mathcal{L}_{\text{cont}}$, a caption loss $\mathcal{L}_{\text{cap}}$, and a sentence loss $\mathcal{L}_{\text{sent}}$ to transfer useful information into T2T.

The app loss $\mathcal{L}_{\text{app}}$ is used in the dynamic feature modulation for app type prediction.
$\mathcal{L}_{\text{app}}$ is a cross-entropy loss, which can be computed as 
$\mathcal{L}_{\text{app}} = - \frac{1}{I}\sum_{i=1}^{I}\log(\mathbf{p}[k_i])$.

The contrastive loss $\mathcal{L}_{\text{cont}}$ is introduced to ensure the distinctiveness of learnable flow pattern prototypes. For the $i$-th sample $\mathbf{T}_i$ with global feature vector $\mathbf{f}_{\text{global}}^{i}$ and true app type label $k_i$, we aim to maximize its similarity to the positive prototypes $\mathbf{P}_{k_i} = \{\mathbf{p}_{k_i,m}\}_{m=1}^M$ and minimize similarity to all negative prototypes from other app types. The similarity to positive prototypes is computed as $z_i^{\text{pos}} = \sum_{m=1}^M \exp(\frac{\mathbf{f}_{\text{global}}^{i} \cdot \mathbf{p}_{k_i,m}}{\tau})$, and to negative prototypes as $z_i^{\text{neg}} = \sum_{k \neq k_i} \sum_{m=1}^M \exp(\frac{\mathbf{f}_{\text{global}}^i \cdot \mathbf{p}_{k,m}}{\tau})$, where $\tau$ is a temperature hyperparameter controlling the similarity distribution. The contrastive loss is then given by 
$\mathcal{L}_{\text{cont}} = -\frac{1}{I} \sum_{i=1}^I \log \frac{z_i^{\text{pos}}}{z_i^{\text{pos}} + z_i^{\text{neg}}}$.

The caption loss \(\mathcal{L}_{\text{cap}}\) is responsible for training the caption decoder. 
This loss is formulated as a cross-entropy loss and compares the predicted word distribution with the ground-truth token \(\mathbf{d}_i = [d^i_1, \dots, d^i_{T_i}]\), where \(T_i\) is the number of tokens in the ground-truth caption. 
The caption loss is formulated as
$\mathcal{L}_{\text{cap}} = - \frac{1}{I} \sum_{i=1}^{I} \sum_{t=1}^{T_i} \log(\mathbf{q}^i_t[d^i_t])$.

The sentence loss $\mathcal{L}_{\text{sent}}$ is employed to align the embeddings of the generated and ground-truth captions~\cite{HMN}. 
For the $i$-th sample $\mathbf{T}_{i}$, we denote $\mathbf{g}_i \in \mathbb{R}^{1\times H'}$ as the embedding of its ground-truth caption generated by Qwen-VL-Max, encoded by SBERT, and mean-pooled and projected into the embedding space $H'$.
$\mathbf{s}_i \in \mathbb{R}^{H'}$ is the embedding of the generated caption.
The sentence loss is calculated as 
$\mathcal{L}_{\text{sent}} = 1 - \frac{1}{I} \sum_{i=1}^{I} \frac{\mathbf{g}_i \cdot \mathbf{s}_i}{\|\mathbf{g}_i\| \|\mathbf{s}_i\|}$.

The overall loss is defined as
\begin{equation}
\mathcal{L}_{\text{overall}} = \lambda_{\text{app}} \mathcal{L}_{\text{app}} + \lambda_{\text{cont}} \mathcal{L}_{\text{cont}} + \lambda_{\text{cap}}  \mathcal{L}_{\text{cap}} + \lambda_{\text{sent}}  \mathcal{L}_{\text{sent}},
\label{eq:total_loss}
\end{equation}
where $\lambda_{\text{app}}, \lambda_{\text{cont}}, \lambda_{\text{cap}}, \lambda_{\text{sent}}$ are weighting coefficients to prioritize different losses during training.

\section{Experimental Evaluation}

\subsection{Evaluation Methodology}
\begin{figure}[t]
    \centering
    \includegraphics[width=0.95\linewidth]{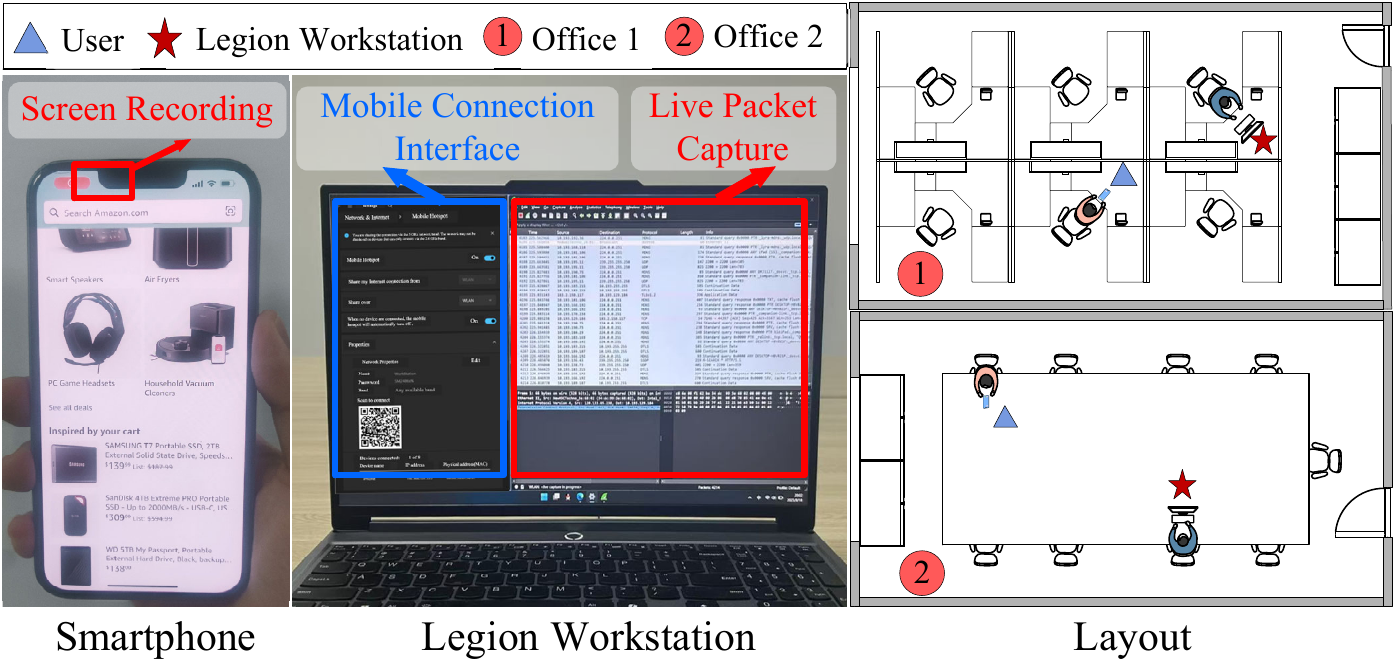}
    \caption{Experimental setups during data collection.}
    \label{fig:setup}
\end{figure}

\begin{table}[t]
\centering
\setlength{\tabcolsep}{2pt}
\caption{Mobile Apps Involved in Data Collection}
\label{tab:applications}
\begin{tabular}{ll}
\hline
\textbf{App Type} & \textbf{Mobile Apps} \\
\hline
Music & Kugou Music, QQ Music, NetEase Cloud Music, Spotify \\
Video & iQIYI, Tencent Video, Youku, YouTube \\
Shopping & Taobao, JD.com, Pinduoduo, Amazon \\
Messaging & QQ, WeChat, Facebook Messenger, WhatsApp \\
Social Media & Baidu Tieba, Rednote, Sina Weibo, Instagram \\
\hline
\end{tabular}
\end{table}

\begin{table}[t]
\centering
\setlength{\tabcolsep}{2pt}   
\caption{Phone Models Used by Volunteers in Two Office Environments}
\label{tab:volunteers}
\begin{tabular}{cc}
\hline
Environment & Phone Models \\
\hline
Office 1 & iPhone 12, iPhone 14, Xiaomi11 Ultra, iQOO Z9 Turbo\\
Office 2 & iPhone 11, iPhone 15 Pro, Huawei P40, REDMI K80 Pro\\
\hline
\end{tabular}
\end{table}

\begin{figure*}[t]
\noindent\centering
\begin{minipage}{0.32\textwidth}
    \centering
    \includegraphics[width=\linewidth]{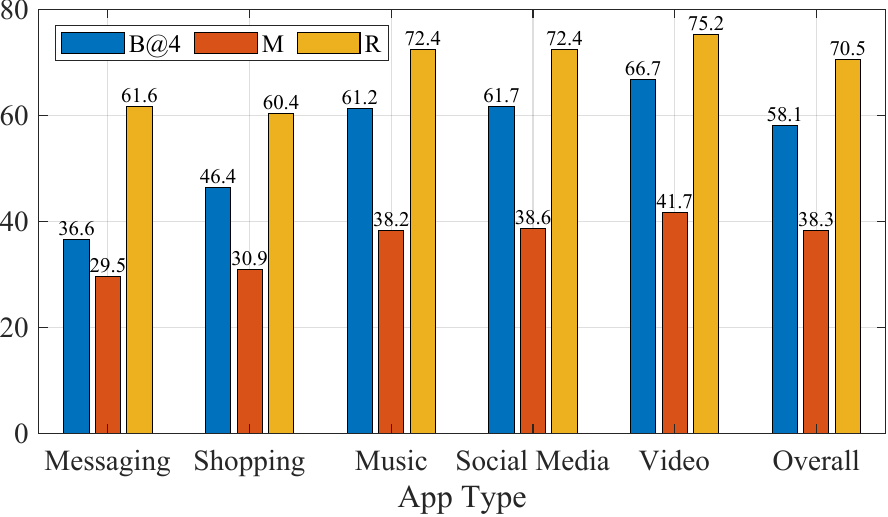}
    \caption{Captioning performance.}
    \label{fig:overall}
\end{minipage}%
\hfill
\begin{minipage}{0.32\textwidth}
    \centering
    \includegraphics[width=\linewidth]{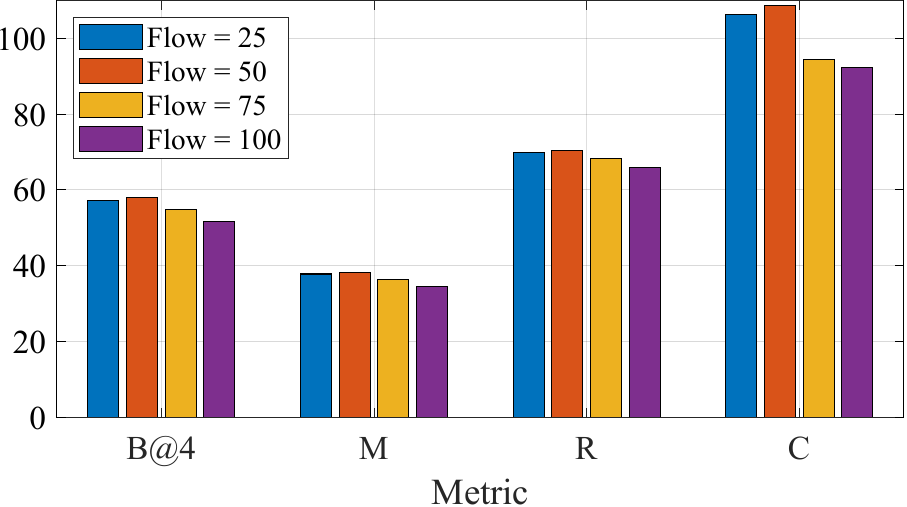}
    \caption{Impact of the flow number.}
    \label{fig:hyper_flow}
\end{minipage}%
\hfill
\begin{minipage}{0.32\textwidth}
    \centering
    \includegraphics[width=\linewidth]{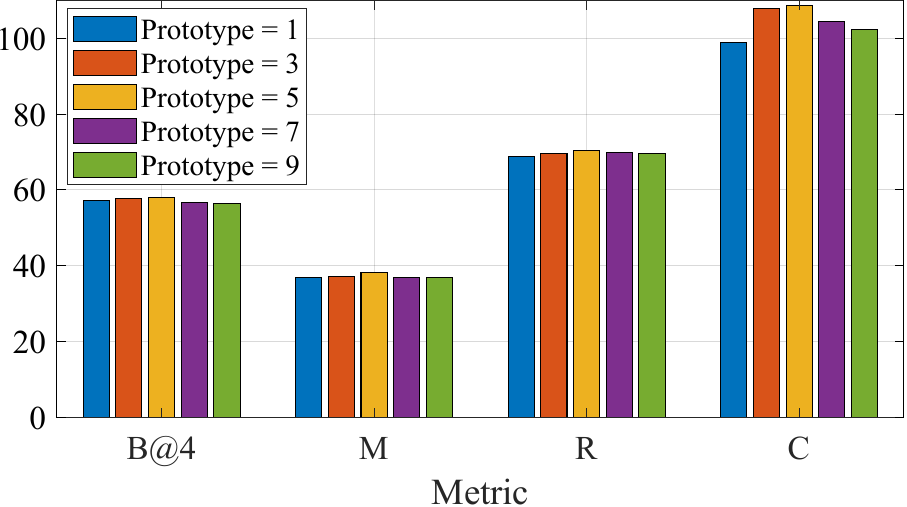}
    \caption{Impact of the prototype number.}
    \label{fig:hyper_prototype}
\end{minipage}%
\end{figure*}
\textbf{Data Collection and Dataset.}
We build a dataset tailored for traffic captioning. 
A Legion Y7000P IRX9 workstation is used as a wireless access point (AP) in two office rooms on our campus, as illustrated in Figure~\ref{fig:setup}. 
8 volunteers are recruited to connect their smartphones to the AP and use the mobile apps listed in Table~\ref{tab:applications}. 
The smartphones used by the volunteers in the two office rooms are summarized in Table~\ref{tab:volunteers}. 
The generated network packets are captured at the workstation in PCAP format using Wireshark. 
When volunteers interact with their smartphones, the screen recording is activated to generate videos of foreground smartphone activities.

The collected traffic is divided into non-overlapping segments lasting 15 seconds each. 
We extract 50 flows from each segment. 
Each flow is represented as a 123-dimensional feature vector~\cite{li2022foap}. 
The collected video is also segmented into 15-second video clips. 
Each clip is aligned with a traffic segment using their timestamps. 
The Qwen-VL-Max model generates 20 captions per video clip, yielding a total of 40,000 traffic--description pairs. 
The dataset is split into 80\% for training, 10\% for validation, and 10\% for testing.


\textbf{Baseline Model.} 
Because no prior work converts network traffic to textual descriptions, we build a baseline model, consisting of a Transformer encoder and an LSTM decoder.
This architecture is widely adopted in image and video captioning approaches~\cite{zhang2024interactive, HMN}.

\textbf{Evaluation Metrics.}
To evaluate the performance of T2T, we use the following four commonly used metrics in natural language generation: BLEU-4, METEOR, ROUGE-L, and CIDEr, which are denoted as B@4, M, R, and C, respectively. 

\textbf{Model Details.}
We set the hidden dimension \(H=512\), the embedding dimension \(H'=256\), the app type embedding dimension \(L=64\), the number of prototypes per app type \(M=5\), the maximum number of network flows \(S=50\), and the temperature for the contrastive loss \(\tau=0.1\).

\subsection{Experimental Results}

\textbf{Captioning Performance.}
First, we analyze the performance of T2T on different app types.
As shown in Fig.~\ref{fig:overall}, T2T achieves the highest BLEU-4, METEOR, and ROUGE-L scores on video apps, slightly lower on social media apps, and the lowest on messaging apps.
This is because frequent video loading or content updating in video and social media apps can trigger abundant burst traffic, which is easier for T2T to recognize. 
In contrast, T2T struggles to characterize subtle flow patterns in messaging apps due to their low-rate and steady traffic caused by text chatting.
Despite that, T2T achieves BLEU-4, METEOR, and ROUGE-L scores of 58.1, 38.3, and 70.5, respectively, suggesting its robustness in generating semantically accurate descriptions.



\begin{figure*}[t]
    \centering
    \includegraphics[width=\textwidth]{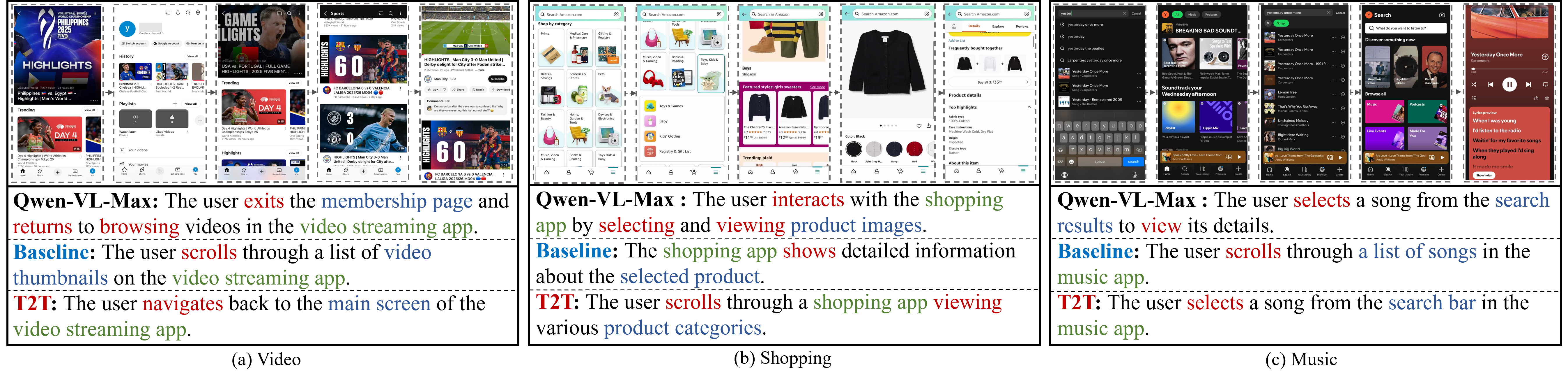}
    \caption{Illustrations of captions generated by Qwen-VL-Max, the baseline model, and T2T. They cover types of mobile apps, including video, shopping, and music. In each caption, verbs are marked in deep red, interface components in blue-gray, and app types in green.}
    \label{fig:generate}
\end{figure*}

\textbf{Hyperparameter Analysis.}
Then, we investigate the impact of the flow number $S$ on T2T’s performance.
As shown in Fig.~\ref{fig:hyper_flow}, T2T achieves optimal performance when $S=50$, with a CIDEr score increase of 2.1\% over 25 flows.
Moreover, it improves BLEU-4, METEOR, ROUGE-L, and CIDEr by 12.3\%, 11.3\%, 6.9\%, and 17.7\%, respectively, over 100 flows.
Accordingly, we set the number of flows to 50.
Furthermore, we analyze the impact of the prototype number. 
As shown in Fig.~\ref{fig:hyper_prototype}, T2T performs best with 5 prototypes, improving the four metrics by 1.7\%, 4.0\%, 2.6\%, and 9.7\% compared to a single prototype.
When the prototype number increases to 7 or 9, T2T also experiences a performance decline.
Specifically, T2T’s CIDEr score decreases by 3.9\% with 7 prototypes and by 5.8\% with 9 prototypes compared to 5 prototypes. 
Thus, we adopt 5 prototypes in T2T for optimal performance.

\begin{table}[t]
\caption{Ablation Study on T2T}
\label{tab:ablation}
\centering
\setlength{\tabcolsep}{5pt}
\begin{tabular}{lcccc}
\toprule
Variants & B@4 & M & R & C \\
\midrule
\#1: w/o DFM \& FPPL (Baseline) & 52.3 & 33.5 & 66.3 & 92.9 \\
\#2: w/o FPPL & 56.8 & 36.1 & 68.7 & 103.8 \\
\#3: w/o DFM & 54.7 & 35.3 & 67.9 & 100.3 \\
\midrule
\#4: T2T (iOS-only dataset) & 62.7 & 39.4 & 72.2 & 110.5 \\
\#5: T2T (Android-only dataset) & 62.5 & 39.6 & 73.4 & 111.7 \\
\midrule
\textbf{T2T (Full)} & \textbf{58.1} & \textbf{38.3} & \textbf{70.5} & \textbf{108.7} \\
\bottomrule
\end{tabular}
\end{table}

\textbf{Ablation Study.}
Next, we evaluate the effectiveness of dynamic feature modulation (DFM), flow pattern prototype learning (FPPL), and platform-specific data distribution in T2T through an ablation study.
As shown in Table~\ref{tab:ablation}, both the second and third variants outperform the baseline, with CIDEr score improvements of 11.7\% and 7.9\%, verifying the effectiveness of DFM and FPPL in generating coherent captions.
Moreover, the second variant shows a 4.5\% lower CIDEr score than T2T, indicating that although DFM provides app type information, it is insufficient to model complex traffic patterns.
Furthermore, the third variant exhibits a larger 7.7\% CIDEr drop compared to T2T, reflecting that FPPL captures flow variations well but lacks sufficient semantic alignment with textual descriptions.
Additionally, the CIDEr score on the complete dataset is slightly lower than on iOS-only and Android-only subsets due to platform-specific traffic differences, yet the performance remains highly comparable, demonstrating the robustness of T2T across diverse platforms.


\textbf{Qualitative Analysis.}
Finally, we compare captions generated by Qwen-VL-Max, the baseline model, and T2T in Fig.~\ref{fig:generate}.
Overall, T2T’s captions are semantically closer to those of Qwen-VL-Max than the baseline model.
For instance, in Fig.~\ref{fig:generate}~(a), Qwen-VL-Max's caption contains ``exits'' and ``returns'', which the baseline omits, while T2T captures similar intent with ``navigates back to the main screen''.  
Meanwhile, the baseline's captions typically contain only a single verb, failing to describe action sequences.
In contrast, T2T generates multiple verbs, thereby providing a more complete transcript of user-smartphone interactions, as shown in Fig.~\ref{fig:generate}~(b).
Even in some cases where T2T's captions have only one verb, they can still convey implicit behavioral information by indicating interface components.
For example, in Fig.~\ref{fig:generate}~(c), T2T generates the phrase ``search bar'', 
suggesting that the user initiates a search operation to find music based on personal preferences.
In summary, T2T can generate semantically rich descriptions from mobile traffic, which is comparable to Qwen-VL-Max.

\section{Conclusion}
In this paper, we present T2T, the first system that generates textual descriptions of smartphone activities from encrypted network traffic, exceeding traditional smartphone activity classification approaches in model scalability and output readability.
T2T addresses two challenges: the semantic gap between traffic features and captions, and the lack of textually annotated traffic data.
We evaluate T2T on 40,000 traffic-description pairs, involving 8 smartphone users and 20 mobile apps, from two real-world environments.
T2T achieves a BLEU-4 score of 58.1, a METEOR score of 38.3, a ROUGE-L score of 70.5, and a CIDEr score of 108.7 on average.
The quantitative and qualitative analyses show that T2T can generate captions that are comparable to the vision–language model.

\section*{Acknowledgments}
This work was supported in part by the National Natural Science Foundation
of China with Grant 62301499 and the Henan Association for Science and Technology with Grant 2025HYTP037.
 
\bibliographystyle{IEEEtran}
\bibliography{references}

\end{document}